# CARAM: A Content-Aware Hybrid PCM/DRAM Main Memory System Framework

Yinjin Fu[0000-0001-9107-1338]

PengCheng Laboratory, Shenzhen 518055, China
`fuyj@pcl.ac.cn`

**Abstract.** The emergence of Phase-Change Memory (PCM) provides opportunities for directly connecting persistent memory to main memory bus. While PCM achieves high read throughput and low standby power, the critical concerns are its poor write performance and limited durability, especially when compared to DRAM. A naturally inspired design is the hybrid memory architecture that fuses DRAM and PCM, so as to exploit the positive aspects of both types of memory. Unfortunately, existing solutions are seriously challenged by the limited main memory size, which is the primary bottleneck of in-memory computing. In this paper, we introduce a novel Content Aware hybrid PCM/DRAM main memory system framework—*CARAM*, which exploits deduplication to improve line sharing with high memory efficiency. CARAM effectively reduces write traffic to hybrid memory by removing unnecessary duplicate line writes. It also substantially extends available free memory space by coalescing redundant lines in hybrid memory, thereby further improving the wear-leveling efficiency of PCM. To obtain high data access performance, we also design a set of acceleration techniques to minimize the overhead caused by extra computation costs. Our experiment results show that CARAM effectively reduces 15%~42% of memory usage and improves I/O bandwidth by 13%~116%, while saving 31%~38% energy consumption, compared to the state-of-the-art of hybrid systems.

## 1 Introduction

The limited main memory capacity has always been a critical issue for multi/many-core systems to meet the needs of concurrent access to working sets. Unfortunately, conventional DRAM is not the ideal storage medium for in-memory computing due to its high power consumption, even though it achieves low access latency. Currently, DRAM-based main memory systems account for about 30%~50% of total power consumption on modern server systems [1][2]. On the other hand, the emerging non-volatile memory (NVM) technologies provide a promising new memory platform to store more data with less power consumption than the expensive-to-build DRAM chips [3]. Phase change memory (PCM), a cutting-edge non-volatile memory technology, is attracting an increasing attention as a promising candidate for next-generation memory [4]. Its desirable characteristics include random access, fast read access, low standby power, and superior scalability. However, there are some crippling limitations that prevent PCM from completely replacing DRAM in future systems, such as low write performance, high power cost of write access, and limited long-term endurance. These drawbacks have led designers toward the adoption of hybrid main memory architectures [5,6,7,8,9,10], which couple the large-capacity PCM with the small-capacity DRAM, in order to combine the best of both memory media.

There are a lot of previous studies on the design of hybrid PCM/DRAM main memory systems. They can be generally divided into two categories: *vertical structure* [5,6,7,8] and *horizontal structure* [9,10,11]. The vertical structure uses DRAM as a cache or buffer for the PCM in order to hide PCM's poor write performance, but it suffers from low space efficiency without mapping the DRAM capacity into memory address space. On the other hand, the horizontal structure manages DRAM and PCM under the same physical address space and allocates their space for page writes, in order to make full use of DRAM capacity with high space efficiency, but it suffers from poor write performance in PCM. Therefore, we focus on a *hybrid* PCM/DRAM main memory design that combines the merits of both vertical and horizontal structures by dividing the DRAM space into two parts: a small part is used for write buffering, while the remaining large part is used for memory address mapping, so as to effectively balance system performance and memory space.

*Deduplication* has become a popular data reduction technique to improve space efficiency by replacing redundant data with references to a unique copy in storage systems[14], due to its excellent ability in removing redundancy with higher throughput than lossless compression techniques[12]. Also, deduplication and compression are orthogonal to each other due to their space savings at different granularity levels, so we can leverage deduplication to complement existing memory compression techniques [16]. Traditional deduplication designs are often deployed in hierarchy memory management with the vertical structure of cache over DRAM[22], DRAM over HDD or SSD[25,26], or NVM over HDD[23, 24, 28] to match the gap of media performance and cost efficiency. In our hybrid PCM/DRAM memory design, we enable deduplication for a hybrid structure that fits the characteristic of storage class memory due to their merits in DRAM-like performance and lower power consumption than DRAM.

In this paper, we present *CARAM*, a content-aware hybrid DRAM/PCM main memory system design framework by leveraging the deduplication technique at the line level. We use DRAM buffering for unique line writes to PCM, and also exploit the available DRAM space for memory address mapping to store deduplication metadata. We also introduce a deduplication-based hybrid memory line write processing to elevate space efficiency by enabling line sharing. Finally, we evaluate the space saving, I/O performance, and power consumption with real-world traces using a simulator that we build for content-aware hybrid memory evaluation.

The main contributions of this paper include:

1) To the best of our knowledge, this is the first study on the architectural design of content-aware hybrid PCM/DRAM main memory system by enabling line sharing to the most extent using deduplication.
2) We present a taxonomy to classify existing work on hybrid main memory into three categories with very different I/O patterns, and motivate our choice of a hybrid structure to balance space saving and system performance.

3) We have implemented a new content-aware hybrid main memory simulator based on DRAMsim2[20] to enable line-level deduplication over hybrid PCM/DRAM memory.
4) Experimental results from our developed simulator implementation of CARAM show that it significantly outperforms state-of-the-art hybrid memory systems in terms of space saving and I/O bandwidth, while achieving almost the same power consumption.

The rest of the paper is organized as follows. We present the background and motivation in Section 2. In Section 3, we propose the system design of our CARAM framework. We describe the experimental evaluation and results in Section 4. In Section 5, we conclude with remarks on future work.

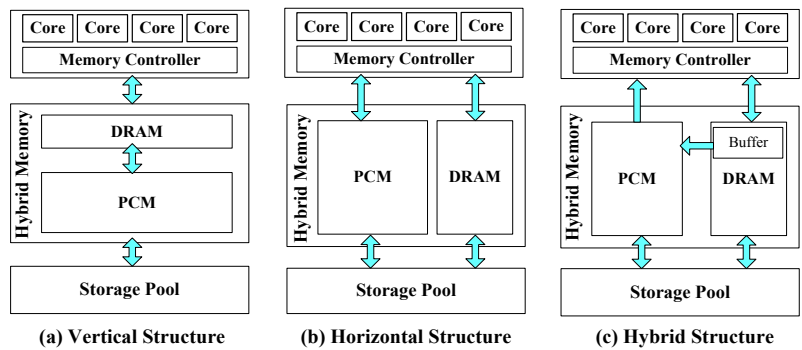

**Fig. 1.** Three kinds of hybrid main memory architecture.

## 2    Background and Motivation

### 2.1    Hybrid Main Memory Architecture

We can generally divide the existing hybrid PCM/DRAM main memory systems into the vertical structure and the horizontal structure. In the vertical structure, as shown in Fig. 1(a), DRAM is used as a cache or buffer for PCM. Zhang and Li [8] consider a hybrid main memory architecture containing a small DRAM-based write buffer to reduce the number of write operations to the PCM partition, and migrates the frequently modified pages from PCM to DRAM so that the lifetime degradation of PCM is alleviated. The authors of [7] study the trade-offs for a main memory system consisting of PCM storage coupled with a small DRAM buffer, which absorbs all incoming traffic from the Last-Level Cache (LLC) to bridge most of the latency gap between DRAM and PCM. In view of this, Ham et al. [6] implement a hybrid memory architecture with disintegrated memory controllers, and the DRAM is used as a cache to provide performance and energy efficiency. Lee et al. [5] introduce a DRAM/PCM hybrid architecture that improves the energy efficiency by employing DRAM as an off-chip cache if the cache hit ratio is high enough to markedly reduce the number of write operations to PCM. All these studies with the vertical structure can hide PCM's poor write performance, but suffer from low space efficiency without mapping the DRAM capacity into memory address space.

In the horizontal structure, DRAM is managed together with PCM under a single physical address space, and address translation is performed via the page table for both DRAM and PCM in Fig. 1(b). PDRAM [9] introduces an efficient operating-system-level page manager for the hybrid memory management to intelligently allocates/ migrates pages across DRAM/PCM in order to minimize the impact of wear leveling on performance. Ramos et al. [11] propose a hybrid memory design that features a hardware-driven rank-based page placement (RaPP) policy, which relies on the memory controller(MC) to monitor access frequency and write intensity in its dynamic ranking of pages, migrates top-ranked popular pages in PCM into DRAM, and moves the unpopular pages in full DRAM to PCM. A hybrid PCM/ DRAM memory management algorithm CLOCK-DWF [10] makes use of the write frequency as well as the recency of write references to accurately estimate future write references. It significantly reduces the number of write operations that occurs on PCM and also increases the lifespan of PCM memory. These researches with the horizontal structure can make full use of the DRAM capacity with high space efficiency, but suffer from poor write performance in PCM.

The work presented in this paper provides a comprehensive design space exploration of DRAM/PCM hybrids from the perspective of space-delay efficiency, i.e., optimizing the combined space and performance behavior of the hybrid system. We propose a novel hybrid structure, as shown in Fig. 1(c), with the merits of both vertical and horizontal structures by dividing the DRAM space into two parts: a small part for write buffering as in the vertical structure, and the remaining large part for memory address mapping as in the horizontal structure. This enables us to effectively balance system performance and memory space. Our hybrid structure is coincident with the popular GB-sized DRAM-based main memory configuration, and it provides a small write buffer to support performance with a few DRAM ranks and the rest DRAM ranks with GB-sized space to support capacity together with many PCM ranks.

### 2.2 Data Deduplication in Hierarchical Memory

Deduplication has been widely applied in designing the memory hierarchy to increase the effective storage size and further improve performance in computing systems. It is a specialized data reduction technique for eliminating duplicate copies of repeating data, and has been widely employed in various high latency scenarios, such as hard-disk-based storage systems [14,15], flash-based solid state drives[18,19] or flash cache [24, 28], and mainly because of its excellent ability in suppressing content redundancy with higher throughput than traditional lossless compression techniques. It works by (1) partitioning a data stream into smaller data objects, (2) representing these objects by their cryptographically secure hash signatures (e.g., MD5 or SHA-1) called fingerprints, (3) replacing the duplicate objects with their fingerprints after the fingerprint index lookup, and (4) transferring or storing only the unique data objects for the purpose of communication and storage efficiency.

However, deduplication in cache and main memory need more considerations on I/O optimization to reduce access latency. Deduplicated LLC uses augmented hashing for fast duplication detection [22], and researchers have used deduplication techniques to store more data in the same cache size, and improved the access performance by reducing the frequency of costly off-chip memory accesses. Memory deduplication has been studied to save DRAM space by identifying and merging identical memory pages [16], and calculating fingerprint with light weighted hashing, such as jhash, CRC32 or SuperFastHash [27]. XLH [17] utilizes cross layer I/O hints in the host's virtual file system to find sharing opportunities earlier without raising the deduplication overhead. SmartMD [25] uses page access information monitored by light weight schemes to improve the efficiency of large page deduplication. UKSM[26] prioritizes different memory regions to accelerate the deduplication process and minimize application penalty. DeWrite [23] enhances the performance and endurance of encrypted NVMs based on a line-level deduplication technique and the synergistic integrations of deduplication and memory encryption.

In hybrid DRAM/PCM main memory architecture, Baek et al. [13] introduce a dual-phase compression scheme, which optimizes a small-size DRAM cache by utilizing a word-level compression algorithm to improve its access latency and effective capacity, and adopts bit-level compression for large-size PCM to further reduce the number of low-performance write accesses, as well as to enhance its lifetime. Note that deduplication and compression are orthogonal to each other due to their merits in space saving at different granularity levels. Our design is the first study on content-aware hybrid PCM/DRAM main memory system by exploiting inline deduplication, so as to significantly improve its space efficiency and I/O performance.

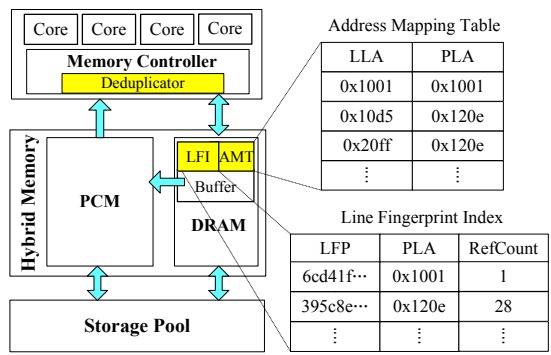

**Fig. 2.** The architecture of CARAM.

## 3 System Design of CARAM

### 3.1 The Overview of System Architecture

Fig. 2 presents our architectural design of CARAM. The on-chip memory controller replays memory requests from the core's LLCs to the DRAM controller or the PCM controller. Line-level deduplication is performed in a deduplicator module, which generates line fingerprints with light-weight hashing SuperFastHash. To support the deduplication process for line fingerprint management and line address mapping in duplicate identification, we store a line fingerprint index (LFI) and an address mapping table (AMT) in a persistent battery-backed DRAM, along with a write buffer for the PCM write accesses to overcome the slow write speed of PCM. We assign all the PCM and the remaining part of DRAM to a single physical memory address space used for the unique line writes after deduplication. Our CARAM design aims to improve space efficiency, power efficiency, and the endurance limit of traditional DRAM/PCM main memory. The whole hybrid main memory is used for page cache to hide disk access latency in the persistent storage pool, and all the unique pages in both DRAM and PCM are managed by page caching algorithms, such as LRU [11] and CLOCK [10].

In our CARAM design, the hybrid main memory serves as a large page cache for the underlying persistent storage pool, and the unique lines from LLC will probably be swapped into disks when the page cache is full. Thus, we need to update its metadata information, such as the line fingerprint index and address mapping table, to limit the size of these in-memory structures. If a swapped/updated line is issued again by a CPU request, it will be regarded as a new line. Accordingly, the deduplication scheme in CARAM can be viewed as an approximate deduplication approach (i.e., a small part of redundant lines may not be deduplicated).

### 3.2  Key Data Structures

In our line-level deduplication in hybrid main memory, there are two hash-map-based data structures: the address mapping table (AMT) and the line fingerprint index (LFI), as shown in Fig. 2. The AMT is an in-memory table that consists of multiple entries, each of which is a key-value pair {logical line address(LLA), physical line address (PLA)}. Each entry requires 4B for storing the LLA and another 4B for PLA. The pair is a many-to-one mapping to support line sharing after deduplication. We need to update the AMT when there are new lines or line updates in hybrid main memory.

The LFI is responsible for the fingerprint management of memory lines. Each of its entries contains a mapping between a line fingerprint (LFP) and a pair: {physical line address (PLA), RefCount}. RefCount presents the corresponding reference count in the hybrid main memory. Each fingerprint is 4B long for SuperFastHash value, while each RefCount is 2B long. Each entry is a one-to-one mapping to record the metadata information of a unique line in main memory at that time. The LFI can be updated when a memory line is renewed or swapped. Unlike the fingerprint index management in the traditional deduplication-based persistent storage, we need to update the LFI in time when a memory page is evicted from the hybrid main memory to the storage pool by the page caching algorithm.

For a hybrid main memory system with 16GB memory space size, we can estimate the sizes of LFI and AMT as about 640MB and 512MB, respectively, for 256B line size and the entry sizes described above. Accordingly, these data structures can be stored in the commonly used 2GB battery-backed DRAM along with the write buffer for PCM writes.

### 3.3  Line Deduplication Processing

As shown in Fig. 3, the line deduplication processing of our CARAM is performed in the deduplicator module of the memory controller. When a line write request on a LPA is issued from the LLC, the corresponding line fingerprint LFP is calculated in CPU using weak hashing SuperFastHash. Then it queries the LFI in DRAM to check whether its line fingerprint exists or not. If yes, it further reads and compares the data in PLA line with the writing line data, if it is duplicate, we can find the mapping of LLA to the same PLA in the AMT, and if it exists in the table, we can deduce that the line write is a duplicate request and drop it; otherwise, the write operation is a duplicate line write, and it updates the AMT for line sharing in hybrid memory. On the other hand, if the line fingerprint LFP does not exist in the LFI or not a duplicate, which means a new line is issued, and it needs to add the new LFP and its metadata into the LFI after the line write is finished in the hybrid main memory. Then it also queries the LLA in the AMT. If it is found, then it updates the AMT for the line update after line edition or page swapping, and deletes the old LFP mapped by LLA in the PFI; Otherwise, it means a new line write is issued, and adds the new LLA into the AMT.

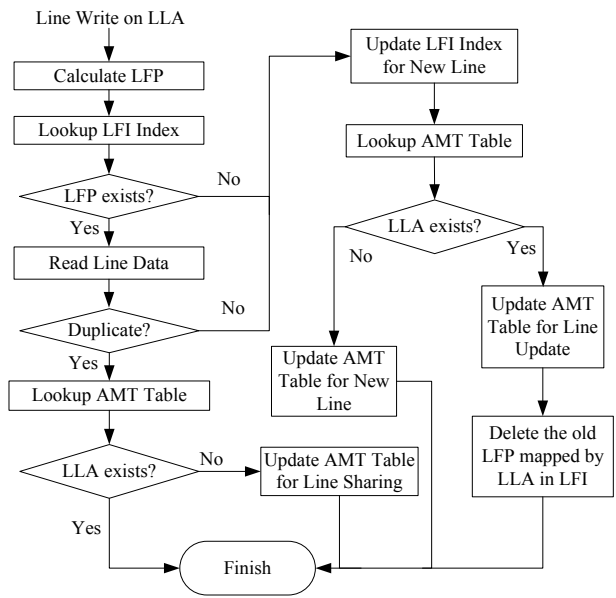

**Fig. 3.** Line deduplication processing in CARAM.

CARAM can efficiently manage the mapping between LLA and PLA in the AMT by updating it for line creation, updates, and sharing. The size of LFI is controlled by the size of hybrid main memory to keep this data structure in the limited size of battery-backed DRAM. For line read request after line deduplication, CARAM can easily find the PLA by looking up the corresponding mapping of its LLA in the AMT.

## 4   Evaluation

In this section, we evaluate hybrid main memory systems in terms of space efficiency, I/O performance, and energy consumption. We compare CARAM with typical existing main memory systems: pure DRAM based memory, pure PCM based memory, and the hybrid DRAM/PCM memory in Fig. 1(c), using our built trace-driven simulator for various experiments. We aim to show the comparison among CARAM and the traditional hybrid main memory without deduplication in terms of space saving, I/O bandwidth, and power consumption.

### 4.1   Experimental Platform and Workload

We conduct experiments on a system running Ubuntu 18.04 with a 3.2GHz quad-core processor, 8GB DRAM, 120GB SSD, and 1TB HDD. We build a trace-driven hybrid memory simulator based on DRAMsim2 [20] for our studies. To support the heterogeneous architecture, we use multiple channels to simulate DRAM and PCM, and Table 1 shows the values of these key parameters in our hybrid memory simulator. We calculate the performance and energy metrics by referring to the energy and performance models used in [5]. There are two channels in the current version: one channel for 8GB PCM, and another channel for 2GB DRAM. As shown in Fig. 4, the new and modified modules are highlighted. We implement the deduplicator module to enable deduplication in hybrid main memory by modifying some components in DRAMsim2 to support simulation for PCM memory. The memory controller receives read/write requests from I/O traces after on-chip cache miss by the simulator interface. It has a queue management module to buffer these requests and their responses in queues. The address map/schedule component will translate the logical address in request into the physical address in DRAM or PCM arrays, and schedule the issuing requests to increase bank usage or page migration between PCM and DRAM. We can easily modify the simulator to support pure DRAM, pure PCM, and the naïve hybrid memory simulation for comparisons.

**Table 1.** Memory parameters in simulator for PCM and DRAM.

| Parameter | | PCM | DRAM |
|---|---|---|---|
| Number of rows | NUM_ROWS | 32768 | 8192 |
| Data bus width | DEVICE_WIDTH (bits) | 8 | 16 |
| Row active time | tRAS (cycles) | 15 | 36 |
| Row address to column address delay | tRCD (cycles) | 5 | 22 |
| Row cycle time | tRC (cycles) | 20 | 96 |
| Row precharge time | tRP (cycles) | 5 | 60 |

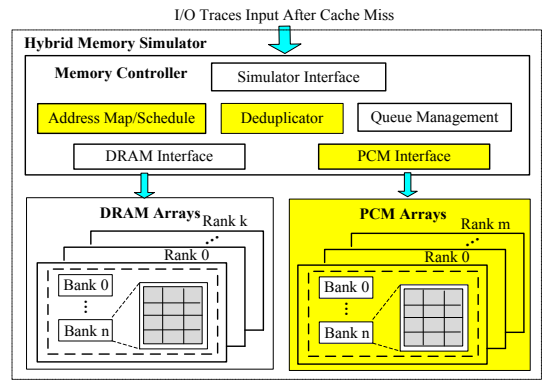

**Fig. 4.** The overview of our hybrid memory simulator.

We feed the simulator with modified I/O traces down-stream of an active cache from four types of application systems: a mail VM server (mail), a web VM server for online course (web-vm), a file server (homes) and a web server for personal pages of users (web-users), in the CS department of FIU for our experiments [21]. The four app-lication workloads were obtained by adding hash calculation time with 1 byte/cycle on each volume as a downstream of a last-level cache, and the summary statistics of one day I/O traces in disk volume size, total and unique read/write accesses are shown in Table 2. It records the same line fingerprint for all 256B lines from every 4KB disk block, and the fingerprint value is the first 4B value of the MD5 value of the disk block.

Table 2. The workload characteristics of I/O traces.

| Workload Type | Volume Size(GB) | Reads (#blocks) | | Write (#blocks) | |
|---|---|---|---|---|---|
| | | *Total* | *Unique* | *Total* | *Unique* |
| mail | 500 | 157012 | 26366 | 212253 | 108664 |
| web-vm | 70 | 42679 | 1341 | 383539 | 146491 |
| homes | 470 | 7368 | 850 | 389559 | 243040 |
| web-users | 10 | 6042 | 143 | 245662 | 172125 |

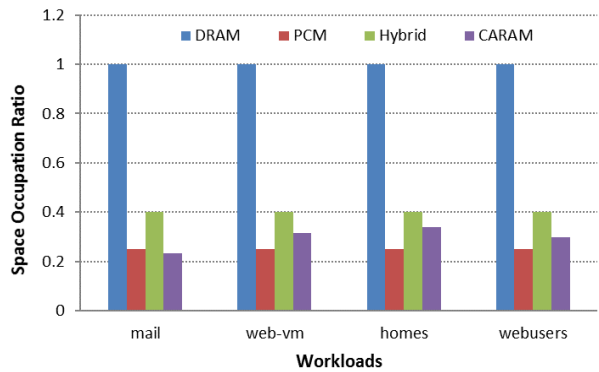

**Fig. 5.** The space efficiency of various main memory architecture.

We assume that the byte price of DRAM is four times that of PCM, and compare CARAM with the three kinds of main memory configuration with the same cost: 4GB DRAM(DRAM), 16GB PCM(PCM), and the hybrid 2GB DRAM+8GB PCM memory (Hybrid). We feed the above I/O trace workloads into our simulator to test the metrics in terms of memory space size, I/O bandwidth, and energy consumption.

### 4.2 Space Efficiency

Limited main memory size is the primary bottleneck for in-memory computing. Different from all the other schemes, CARAM employs line-level deduplication to improve the space efficiency of hybrid main memory with low system overhead. As deduplication is orthogonal to compression, we can save more memory space if we further compress the unique lines in hybrid memory after deduplication. However, our workloads are collected I/O traces which only contain the hash signatures of the lines without real data content for further compression. We use the space occupation ratio as a metric for space saving, defined as the actual physical memory space overhead divided by the total memory capacity. Here, we assume the space occupation ratio of DRAM is 1 in all four applications. As shown in Fig. 5, the ratio of PCM is very low, and it has only a quarter of DRAM's value due to its low unit price. The ratio of the naïve hybrid memory is more than 0.4, while CARAM can improve it to approach or even better than that of PCM, by saving 15%~42% memory space of hybrid main memory systems via deduplication.

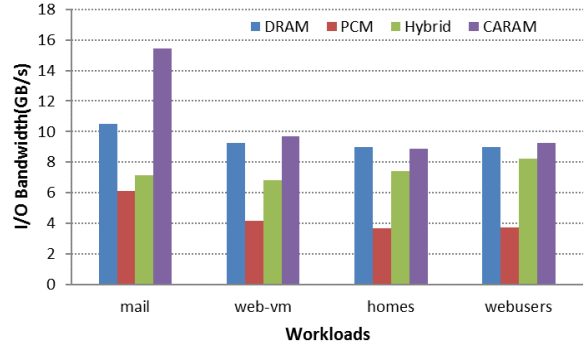

**Fig. 6.** The I/O performance of different memory frameworks.

### 4.3 I/O Performance

Data intensive applications can be accelerated by executing computing tasks purely in high-performance main memory rather than a combination of memory and disk. DRAM-based main memory can achieve the highest I/O performance with some drawbacks in static power and volatile property. PCM-based main memory is a lower performance non-volatile memory which has larger capacity and lower power consumption than DRAM. Hybrid main memory can effectively balance the I/O performance and power consumption of the above two kinds of main memory. CARAM enables deduplication in hybrid main memory, and it can significantly reduce line writes to enhance I/O performance through line sharing. Fig. 6 shows that CARAM can achieve higher performance range from 13% to 116% than the naïve hybrid main memory system, since it performs a large number of low-overhead metadata updates only instead of duplicate line writes. Also, it performs the best under the mail workload mainly due to its lowest space occupation ratio.

### 4.4 Energy Consumption

The energy consumption of main memory is becoming a dominant portion of the total system energy due to the increasing memory capacity driven by data intensive applications. DRAM is volatile and needs the periodical power-hungry refresh processes to keep its information readable. PCM's non-volatility obviates refresh with very low power consumption, but it has a higher power cost of write accesses compared to DRAM. The hybrid PCM and DRAM memory architecture can achieve a significant energy saving at negligible performance overhead over comparable DRAM organization. CARAM can improve the energy efficiency of hybrid main memory by exploiting line-level deduplication to perform several metadata reads and updates instead of duplicate writes in PCM. We evaluate the energy consumption of the four memory architectures with the I/O traces first, but their values are almost the same under the four different workloads due to the domination of idle time. To differentiate these schemes, we stress-test their energy consumption by continuously issuing the read or write line request without the greatest common idle intervals. Results in Fig. 7 show that our CARAM can save 31%~38% energy consumption than that of the naïve hybrid main memory, since it can significantly reduce the number of write operations in main memory.

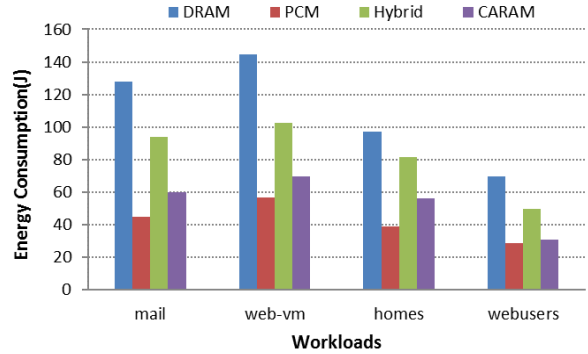

**Fig. 7.** The comparison of energy consumption on the main memory systems.

## 5     Conclusions

In this paper, we present a content aware hybrid DRAM/PCM main memory system, called CARAM, by exploiting line sharing with deduplication technique, and implement it in a trace-driven hybrid memory simulator based on DRAMsim2. Specifically, we introduce line-level deduplication processing of write access in our hybrid structure to balance space saving and system performance. Evaluation results show that CARAM constantly outperforms the existing hybrid memory systems in terms of space saving, I/O bandwidth, and power consumption. We will study the combination of deduplication and memory compression running real data as a direction of future work.

### Acknowledgments

This research was supported by the NSF-Jiangsu grant BK20191327. We would like to thank Prof. Patrick P.C. Lee and Dr. Yang Wu for their help on the initial design of the system.